\documentclass[twocolumn,aps,pra,superscriptaddress,showpacs,tightenlines]{revtex4-2}
\usepackage{amsmath}
\usepackage{amsfonts}
\usepackage{graphicx}
\usepackage{epsfig}
\usepackage{units}
\usepackage{color}
\usepackage[colorlinks,linkcolor=blue,anchorcolor=blue,citecolor=blue,urlcolor=blue]{hyperref}
\usepackage{multirow}

\usepackage{booktabs}
\begin{document}
\title{Dynamical emission of phonon pairs in optomechanical systems}

\author{Fen Zou}
\affiliation{Beijing Computational Science Research Center, Beijing 100193, China}

\author{Jie-Qiao Liao}
\email{jqliao@hunnu.edu.cn}
\affiliation{Key Laboratory of Low-Dimensional Quantum Structures and Quantum Control of
Ministry of Education, Key Laboratory for Matter Microstructure and Function of Hunan Province, Department of Physics, Hunan Normal University, Changsha 410081, China}
\affiliation{Synergetic Innovation Center for Quantum Effects and Applications, Hunan Normal University, Changsha 410081, China}

\author{Yong Li}
\email{liyong@csrc.ac.cn}
\affiliation{Beijing Computational Science Research Center, Beijing 100193, China}
\affiliation{Synergetic Innovation Center for Quantum Effects and Applications, Hunan Normal University, Changsha 410081, China}

\date{\today}

\begin{abstract}
Multiphonon state plays an important role in quantum information processing and quantum metrology. Here we propose a scheme to realize dynamical emission of phonon pairs based on the technique of stimulated Raman adiabatic passage in a single cavity optomechanical system, where the optical cavity is driven by two Gaussian pulse lasers. By exploring quantum trajectories of the state populations and the average phonon number, we find that the dynamical phonon-pair emission can be realized under the appropriate parameter conditions and is tunable by controlling the time interval between the consecutive pulses of pump lasers. In particular, the numerical results for the standard and generalized second-order correlation functions of the mechanical mode show that the system can behave as an antibunched phonon-pair emitter. Our proposal can be extended to achieve an antibunched $n$-phonon emitter, which has potential applications for on-chip quantum communications.
\end{abstract}
\maketitle

\section{Introduction}
The realization of the nonclassical states has become an interesting and important research topic in quantum information science, with potential applications in quantum communication~\cite{kimble2008Quantum}, quantum metrology~\cite{giovannetti2006Quantum}, quantum lithography~\cite{dangelo2001TwoPhoton}, quantum spectroscopy~\cite{lopezcarreno2015Exciting,dorfman2016Nonlinear}, and quantum biology~\cite{denk1990Twophoton,horton2013Vivo}. Recently, the generation of $n$-quanta states has been studied theoretically in multilevel atomic systems~\cite{dousse2010Ultrabright,Yasutomo2011Spontaneous,koshino2013Implementation,callsen2013Steering,
muller2014Ondemand,munoz2015Enhanced,chang2016Deterministic,hargart2016Cavityenhanced,sanchez-burillo2016Full,dong2019Multiphonon}, Rydberg atomic ensembles~\cite{bienias2014Scattering,maghrebi2015Coulomb}, cavity quantum electrodynamics (QED) systems~\cite{munoz2014Emitters,strekalov2014Bundle,munoz2018Filtering,bin2020Phonon,bin2021ParitySymmetryProtected,deng2021Motional,
cosacchi2021suitability,Zhu2017Collective,Bin2018Two,zou2020Multiphoton}, circuit QED systems~\cite{ma2021Antibunched}, waveguide QED systems~\cite{gonzalez-tudela2015Deterministic,douglas2016Photon,gonzalez-tudela2017Efficient}, Kerr cavity systems~\cite{Miranowicz2013Two,Huang2018Nonreciprocal}, and cavity optomechanical systems~\cite{qin2019Emission}. In particular, emitter of $n$-photon bundles, releasing their energy in the bundle of $n$ photons, was first proposed by C. S\'{a}nchez Mu\~{n}oz et al. in a cavity QED system~\cite{munoz2014Emitters}.

Subsequently, a series of schemes on the $n$-photon and $n$-phonon bundle emissions have been proposed in a variety of quantum systems, e.g., cavity~\cite{munoz2018Filtering,bin2020Phonon,bin2021ParitySymmetryProtected,deng2021Motional,
cosacchi2021suitability} and circuit QED systems~\cite{ma2021Antibunched}. The antibunched $n$-photon and $n$-phonon bundle emissions can be used to realize multiphoton and multiphonon sources~\cite{satzinger2018Quantum,chu2018Creation}, respectively. However, since the high-order process of the single-photon (single-phonon) transition is generally very weak, the experimental realization of multiphoton (multiphonon) state is still a challenge.

In this work, we propose a scheme for implementing dynamical phonon-pair emission in a cavity optomechanical system composed of an optical cavity and a mechanical resonator~\cite{Kippenberg2008Cavity,Aspelmeyer2012Quantum,Aspelmeyer2014Cavity,Law1995Interaction,Liao2012Spectrum,Liao2013Correlated}, where the optical cavity is driven by two Gaussian pulse lasers. Under the appropriate parameter conditions, the dimensions of the Hilbert space of the cavity and mechanical modes are truncated up to $1$ and $2$ excitations, respectively. Based on the technique of stimulated Raman adiabatic passage (STIRAP)~\cite{gaubatz1990Population,bergmann1998Coherent,vitanov2017Stimulated}, the population transfer between zero-phonon state and two-phonon state can be realized in the absence of dissipation in the system. In the presence of dissipation, we find that the dynamical phonon-pair emission can be observed by analyzing quantum trajectories of the state populations and the average phonon number in the system. In addition, we investigate quantum statistics of the dynamical phonon-pair emission by numerically calculating the standard and generalized second-order correlation functions in the mechanical mode. It can be found that the system behaves as an antibunched phonon-pair emitter when the time interval $T$ between the consecutive pulses of pulse lasers is much larger than the mechanical lifetime $1/\gamma_{m}$, i.e., $\gamma_{m}T\gg1$. Particularly, compared to the previous $n$-phonon bundle emission~\cite{bin2020Phonon,deng2021Motional}, the time interval of the dynamical phonon-pair emission can be tuned by adjusting the time interval between the consecutive pulses. Our work opens up a route to achieve an antibunched phonon-pair emitter, which could be useful for quantum information processing and for medical applications.

The rest of this paper is organized as follows. In Sec.~\ref{modelsec}, we introduce the physical model and present the Hamiltonian of the system. In Sec.~\ref{effHamtwophonsec}, we derive an effective Hamiltonian of the system in a finite-dimensional Hilbert space and analyze the generation of the two-phonon state. In Sec.~\ref{tpbesec}, we study the dynamical emission of phonon pairs by analyzing quantum trajectories of the state populations and the average phonon number in the system. We also investigate the statistical properties of the dynamical phonon-pair emission by numerically calculating the standard and generalized second-order correlation functions in the mechanical mode. Finally, we present some discussions on the experimental parameters and conclude this work in Sec.~\ref{conclusion}.

%%%%%%%%%%%%%%%%%%%%%
\begin{figure}
\center
\includegraphics[width=0.4 \textwidth]{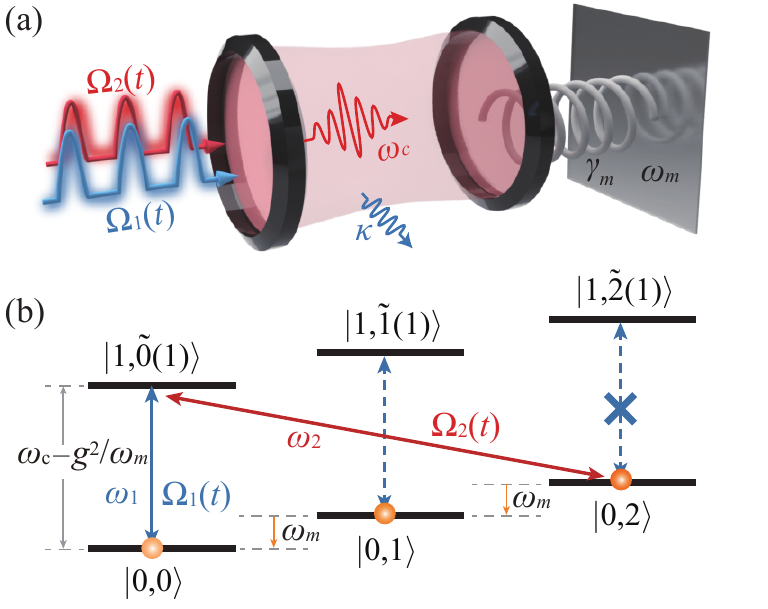}
\caption{(a) Schematic diagram of a cavity optomechanical system consisting of a single cavity mode coupled to a mechanical mode via radiation-pressure interaction. The optical cavity is driven by two Gaussian pulse driving fields $\Omega_{1}(t)$ and $\Omega_{2}(t)$. (b) The resonant transition of the effective Hamiltonian $H_{\text{eff}}^{(2)}$ at the coupling strength $g/\omega_{m}\approx0.765$, where there is no transition between $\vert0,2\rangle$ and $\vert1,\tilde{2}(1)\rangle$. Other parameters are $\Delta_{1}\equiv\omega_{1}-\omega_{c}=-g^{2}/\omega_{m}$ and $\Delta_{2}\equiv\omega_{2}-\omega_{c}=-g^{2}/\omega_{m}-2\omega_{m}$.}
\label{Fig1}
\end{figure}
%%%%%%%%%%%%%%%%%%%%%%%
\section{Model and Hamiltonian \label{modelsec}}
As schematically shown in Fig.~\ref{Fig1}(a), we consider a cavity optomechanical system which consists of a single cavity mode coupled to a mechanical mode via radiation-pressure interaction. The cavity is driven by two Gaussian pulse driving fields with corresponding driving carrier frequencies $\omega_{1}$ and $\omega_{2}$, where each driving field is composed of a series of consecutive Gaussian wave packets. The Hamiltonian of the system reads ($\hbar=1$)
\begin{align}\label{Ham}
H&=\omega_{c}a^{\dagger}a+\omega_{m}b^{\dagger}b-ga^{\dagger}a(b^{\dagger}+b)\nonumber\\
&\quad+[\Omega_{1}(t)a^{\dagger}e^{-i\omega_{1}t}+\Omega_{2}(t)a^{\dagger}e^{-i\omega_{2}t}+\text{H.c.}]
\end{align}
with the time-dependent amplitudes of the driving fields
\begin{equation}
\Omega_{i}(t)=\Omega_{0}\sum_{k=0}^{\infty}\exp\left[-\frac{(t-t_{i}-kT)^{2}}{2\sigma^{2}}\right],\quad i=1,2.
\end{equation}
Here $a^{\dagger}$ ($a$) is the creation (annihilation) operator of the cavity mode with resonance frequency $\omega_{c}$, $b^{\dagger}$  ($b$) is the creation (annihilation) operator of the mechanical mode with resonance frequency $\omega_{m}$, and $g$ is the single-photon optomechanical coupling strength. The parameters $\Omega_{0}$ and $\sqrt{2}\sigma$ are, respectively, the amplitude and duration of the Gaussian wave packets~\cite{Hovsepyan2015Excitations}. $t_{1}+kT$ ($t_{2}+kT$) corresponds to the time instant when the pulse $\Omega_{1}(t)$ [$\Omega_{2}(t)$] reaches its maximum value, where $k$ is an integer and $T$ is the time interval of consecutive Gaussian wave packets.

In the rotating frame with respect to $\omega_{c}$, Hamiltonian~(\ref{Ham}) becomes
\begin{align}\label{Hamr}
H_{r}&=\omega_{m}b^{\dagger}b-ga^{\dagger}a(b^{\dagger}+b)+[\Omega_{1}(t)a^{\dagger}e^{-i\Delta_{1}t}\nonumber\\
&\quad+\Omega_{2}(t)a^{\dagger}e^{-i\Delta_{2}t}+\text{H.c.}],
\end{align}
where $\Delta_{1}=\omega_{1}-\omega_{c}$ and $\Delta_{2}=\omega_{2}-\omega_{c}$ are detunings between the carrier frequencies of two pulse driving fields and the cavity frequency, respectively.

By introducing a conditional displacement operator $D(\beta a^{\dagger}a)=\text{exp}[\beta a^{\dagger}a(b^{\dagger}-b)]$ with $\beta=g/\omega_{m}$, the first two terms $H_{s}=\omega_{m}b^{\dagger}b-ga^{\dagger}a(b^{\dagger}+b)$ of Eq.~(\ref{Hamr}) can be diagonalized as
\begin{equation}
H_{s}=\sum_{n,m=0}^{\infty}E_{n,m}\vert n,\tilde{m}(n)\rangle\langle n,\tilde{m}(n)\vert,
\end{equation}
where the eigenstates of $H_{s}$ are $\vert n,\tilde{m}(n)\rangle=\vert n\rangle_{a}\otimes\vert \tilde{m}(n)\rangle_{b}=\vert n\rangle_{a}\otimes D(n\beta)\vert m\rangle_{b}$, and the corresponding eigenvalues are $E_{n,m}=m\omega_{m}-g^{2}n^{2}/\omega_{m}$. Here $\vert n\rangle_{a}$ ($n=0,1,2,\ldots$) are number states of the cavity mode, $\vert m\rangle_{b}$ ($m=0,1,2,\ldots$) are number states of the mechanical mode, and $\vert \tilde{m}(n)\rangle_{b}$ are the $n$-photon displaced number states of the mechanical mode. In particular, when $n=0$, we have $\vert 0,\tilde{m}(0)\rangle=\vert 0,m\rangle$.

By using the eigenbasis of the Hamiltonian $H_{s}$, Hamiltonian~(\ref{Hamr}) can be further written as
\begin{align}\label{HamrEig}
H_{r}&=H_{s}+\sum_{n,m,q=0}^{\infty}\{A_{m,q}^{(n)}[\Omega_{1}(t)e^{-i\Delta_{1}t}+\Omega_{2}(t)e^{-i\Delta_{2}t}] \nonumber\\
&\quad\times\vert n,\tilde{m}(n)\rangle\langle n-1,\tilde{q}(n-1)\vert+\text{H.c.}\},
\end{align}
where we introduce the coefficients $A_{m,q}^{(n)}=\sqrt{n}\,_{b}\langle m\vert D(-\beta)\vert q\rangle_{b}$. The coefficients can be calculated by~\cite{deoliveira1990Properties}
\begin{align}
A_{m,q}^{(n)}=\begin{cases}
\sqrt{n}\sqrt{\frac{m!}{q!}}e^{-\frac{\beta^{2}}{2}}\beta^{q-m}L_{m}^{q-m}(\beta^{2}), & m<q,\\
\sqrt{n}\sqrt{\frac{q!}{m!}}e^{-\frac{\beta^{2}}{2}}(-\beta)^{m-q}L_{q}^{m-q}(\beta^{2}), & m\geq q,
\end{cases}
\end{align}
where $L_{m}^{q}(x)$ are the associated Laguerre polynomials. In the rotating frame with respect to $H_{s}$, the Hamiltonian (\ref{HamrEig}) is transformed as
\begin{align}\label{HamI}
H_{I}&=\sum_{n,m,q=0}^{\infty}\{A_{m,q}^{(n)}[\Omega_{1}(t)e^{i(\delta_{n,m,q}-\Delta_{1})t}+\Omega_{2}(t) \nonumber\\
&\quad\times e^{i(\delta_{n,m,q}-\Delta_{2})t}]\vert n,\tilde{m}(n)\rangle\langle n-1,\tilde{q}(n-1)\vert+\text{H.c.}\},
\end{align}
where the variable $\delta_{n,m,q}=E_{n,m}-E_{n-1,q}=(m-q)\omega_{m}-g^{2}(2n-1)/\omega_{m}$.

\section{Effective Hamiltonian and generation of the two-phonon state \label{effHamtwophonsec}}
In this section, we will derive the effective Hamiltonian for $H_{I}$ in a finite-dimensional Hilbert space and analyze the generation of a two-phonon state.

\subsection{Effective Hamiltonian in a confined Hilbert space \label{effHamsec}}
To analyze the generation of $N$-phonon states, we derive the effective Hamiltonian for $H_{I}$ in a finite-dimensional Hilbert space. Under the condition of the resolved sideband (i.e., the cavity-field decay rate $\kappa$ is much smaller than the mechanical frequency $\omega_{m}$), we choose the driving carrier frequencies $\omega_{1}$ and $\omega_{2}$ to satisfy the resonance transitions of $\vert 0,m\rangle\overset{\Omega_{1}(t)}{\longleftrightarrow}\vert 1,\tilde{m}(1)\rangle$ and $\vert 0,m+2\rangle\overset{\Omega_{2}(t)}{\longleftrightarrow}\vert 1,\tilde{m}(1)\rangle$, respectively [see Fig.~\ref{Fig1}(b)]. Hence, the two driving detunings are $\Delta_{1}=-g^{2}/\omega_{m}$ and $\Delta_{2}=-g^{2}/\omega_{m}-2\omega_{m}$. In this circumstance, Hamiltonian~(\ref{HamI}) can be broken down into two parts:
\begin{equation}
H_{I}=\tilde{H}_{I}+H_{I}^{\prime},
\end{equation}
where $\tilde{H}_{I}$ denotes the resonant transitions
\begin{align}\label{HItilde}
\tilde{H}_{I}&=\sum_{m=0}^{\infty}[\Omega_{1}(t)A_{m,m}^{(1)}\vert1,\tilde{m}(1)\rangle\langle 0,m\vert \nonumber\\
&\quad+\Omega_{2}(t)A_{m,m+2}^{(1)}\vert1,\tilde{m}(1)\rangle\langle 0,m+2\vert]+\text{H.c.},
\end{align}
and $H_{I}^{\prime}$ corresponds to the off-resonant transitions
\begin{align}\label{HIprime}
H_{I}^{\prime}&=\sum_{n,m,q=0}^{\infty}\!^{\prime}\{A_{m,q}^{(n)}[\Omega_{1}(t)e^{i\delta_{n,m,q}^{(1)}t}+\Omega_{2}(t)e^{i\delta_{n,m,q}^{(2)}t}] \nonumber\\
&\quad\times\vert n,\tilde{m}(n)\rangle\langle n-1,\tilde{q}(n-1)\vert+\text{H.c.}\}.
\end{align}
Here the primed summation in Eq.~(\ref{HIprime}) excludes those terms of the Hamiltonian $\tilde{H}_{I}$, and the off-resonance detunings $\delta_{n,m,q}^{(i)}=\delta_{n,m,q}-\Delta_{i}$ are given by
\begin{subequations}
\begin{align}
\delta_{n,m,q}^{(1)}&=(m-q)\omega_{m}-\frac{2g^{2}}{\omega_{m}}(n-1),\label{deltai1}\\
\delta_{n,m,q}^{(2)}&=(m-q+2)\omega_{m}-\frac{2g^{2}}{\omega_{m}}(n-1).\label{deltai2}
\end{align}
\end{subequations}
Here $q\neq m$ in Eq.~(\ref{deltai1}) and $q\neq m+2$ in Eq.~(\ref{deltai2}) for $n=1$.

\begin{figure}
\center
\includegraphics[width=0.46 \textwidth]{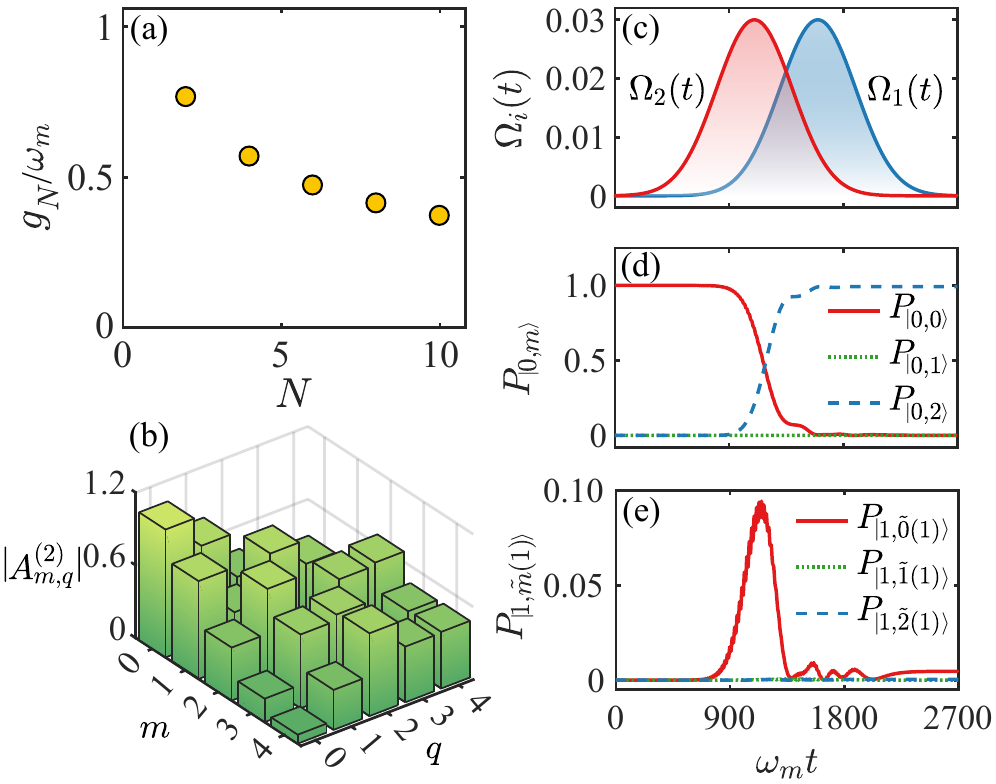}
\caption{(a) The ratio $g_{N}/\omega_{m}$ of the minimal positive value $g_{N}$ satisfying Eq.~(\ref{Ann}) over the mechanical frequency $\omega_{m}$ as a function of the index $N$. (b) The coefficients $\vert A_{m,q}^{(2)}\vert$ as functions of the indexs $m$ and $q$ at $g/\omega_{m}\approx0.765$. (c)-(e) The first Gaussian wave packets
of two pulse lasers $\Omega_{i}(t)$ $(i=1,2)$, the zero-photon state populations $P_{\vert0,m\rangle}=\vert\langle0,m\vert\psi(t)\rangle\vert^{2}$, and the single-photon state populations $P_{\vert1,\tilde{m}(1)\rangle}=\vert\langle1,\tilde{m}(1)\vert\psi(t)\rangle\vert^{2}$ as functions of the time $\omega_{m}t$. Here $\vert\psi(t)\rangle$ is the state of the system at time $t$ in the absence of dissipation. Other parameters are $\Omega_{0}/\omega_{m}=0.03$, $\omega_{m}\sigma=300$, $\omega_{m}t_{1}=1600$, $\omega_{m}t_{2}=1100$, $g/\omega_{m}\approx0.765$, $\Delta_{1}=-g^{2}/\omega_{m}$, and $\Delta_{2}=-g^{2}/\omega_{m}-2\omega_{m}$.}
\label{Fig2}
\end{figure}

In order to neglect the off-resonant transition part $H_{I}^{\prime}$, the parameter conditions $\vert\delta_{n,m,q}^{(i)}\vert\gg \vert A_{m,q}^{(n)}\vert\Omega_{0}$ $(i=1,2)$ should be satisfied. In particular, in order to ignore $H_{I}^{\prime}$, we need to block the transitions from one-photon states $\vert1,\tilde{m}(1)\rangle$ to two-photon states $\vert2,\tilde{m}(2)\rangle$, and this requires $\vert\delta_{2,m,q}^{(i)}\vert\gg \vert A_{m,q}^{(2)}\vert\Omega_{0}$. Since the coefficients $\vert A_{m,q}^{(2)}\vert\lesssim1$, as shown in Fig.~\ref{Fig2}(b), the parameter condition is~\cite{Xu2013Dark}
\begin{equation}\label{Paracond}
\Omega_{0}\ll\left|\frac{2g^{2}}{\omega_{m}}-K\omega_{m}\right|
\end{equation}
with $K$ being the nearest integer to $2(g/\omega_{m})^{2}$. Hence, when the parameter condition of Eq.~(\ref{Paracond}) is satisfied, the Hamiltonian $H_{I}$ can be approximately reduced to $\tilde{H}_{I}$, which describes the resonant transitions between zero-photon states $\vert 0,m\rangle$ and one-photon states $\vert 1,\tilde{m}(1)\rangle$.

%%%%%%%%%%%%%%%%%%%%%
%%%%%%%%%%%%%%%%%%%%%%%

In addition, the dimension of the Hilbert space of the mechanical mode can also be approximately truncated up to $m=N$ by choosing the single-photon optomechanical coupling strength $g=g_{N}$, where $g_{N}$ is the minimal positive value satisfying the following equation:
\begin{equation}\label{Ann}
A_{N,N}^{(1)}=\exp\left(-\frac{g_{N}^{2}}{2\omega_{m}^{2}}\right)L_{N}^{0}\left(\frac{g_{N}^{2}}{\omega_{m}^{2}}\right)=0,
\end{equation}
with $N$ being a positive even number. It can be seen from Eq.~(\ref{Ann}) that the transition matrix element $\Omega_{1}(t)A_{N,N}^{(1)}$ from the state $\vert 0,N\rangle$ to $\vert 1,\tilde{N}(1)\rangle$ is zero. In Fig.~\ref{Fig2}(a), we show the dependence of the ratio $g_{N}/\omega_{m}$ of the minimal positive value $g_{N}$ satisfying Eq.~(\ref{Ann}) over the mechanical frequency $\omega_{m}$ on the index $N$. The result shows that the coupling strength $g_{N}$ decreases as the truncation dimension $N$ increases.

When the parameter conditions of Eqs.~(\ref{Paracond}) and~(\ref{Ann}) are satisfied, the effective Hamiltonian of $H_{I}$ can be obtained as
\begin{align}\label{Heff}
H_{\text{eff}}^{(N)}&=\sum_{m=0}^{N-1}\Omega_{1}(t)A_{m,m}^{(1)}\vert1,\tilde{m}(1)\rangle\langle 0,m\vert+\sum_{m=0}^{N-2}\Omega_{2}(t) \nonumber\\
&\quad\times A_{m,m+2}^{(1)}\vert1,\tilde{m}(1)\rangle\langle 0,m+2\vert+\text{H.c.},
\end{align}
where $H_{\text{eff}}^{(N)}$ describes the resonant transitions between zero-photon states $\vert 0,m\rangle$ and one-photon states $\vert 1,\tilde{m}(1)\rangle$ with the phonon number $m\leq N$.

\subsection{Generation of the two-phonon state \label{twophonsec}}
We now analyze the generation of the two-phonon state based on technique of STIRAP. When considering the single-photon optomechanical coupling strength $g/\omega_{m}=g_{2}/\omega_{m}\approx0.765$, the dimension of the Hilbert space of the mechanical mode can be truncated up to $N=2$, the effective Hamiltonian can then be expressed as
\begin{align}\label{Heff2}
H_{\text{eff}}^{(2)}&=\Omega_{1}(t)[A_{0,0}^{(1)}\vert1,\tilde{0}(1)\rangle\langle 0,0\vert+A_{1,1}^{(1)}\vert1,\tilde{1}(1)\rangle\langle 0,1\vert] \nonumber\\
&\quad+\Omega_{2}(t)A_{0,2}^{(1)}\vert1,\tilde{0}(1)\rangle\langle 0,2\vert+\text{H.c.}.
\end{align}
In Fig.~\ref{Fig1}(b), we show the resonant transition $\vert0,0\rangle\overset{\Omega_{1}(t)}{\longleftrightarrow}\vert1,\tilde{0}(1)\rangle
\overset{\Omega_{2}(t)}{\longleftrightarrow}\vert0,2\rangle$ of the effective Hamiltonian $H_{\text{eff}}^{(2)}$. When the initial state of the system is $\vert0,0\rangle$, the transition $\vert0,1\rangle\overset{\Omega_{1}(t)}{\longleftrightarrow}\vert1,\tilde{1}(1)\rangle$ is negligible in the absence of dissipation.

For the effective Hamiltonian $H_{\text{eff}}^{(2)}$, the matrix form can be expressed as
\begin{equation}\label{Meff2}
H_{\text{eff}}^{(2)}=M_{33}\oplus M_{22}
\end{equation}
with
\begin{equation}\label{M33}
M_{33}=\left(
\begin{array}{ccc}
0 & \Omega_{1}\left(t\right)A_{0,0}^{\left(1\right)} & 0  \\
\Omega_{1}\left(t\right)A_{0,0}^{\left(1\right)} & 0 & \Omega_{2}\left(t\right)A_{0,2}^{\left(1\right)}   \\
0 & \Omega_{2}\left(t\right)A_{0,2}^{\left(1\right)} & 0
\end{array}
\right),
\end{equation}
and
\begin{equation}\label{M22}
M_{22}=\left(
\begin{array}{cc}
0 & \Omega_{1}\left(t\right)A_{1,1}^{\left(1\right)}\\
\Omega_{1}\left(t\right)A_{1,1}^{\left(1\right)} & 0
\end{array}
\right).
\end{equation}
Here the symbol ``$\oplus$" denotes direct sum of the matrix, and the matrix (\ref{Meff2}) is defined based on the basis states $\vert0,0\rangle=(1,0,0,0,0)^{\text{T}}$, $\vert1,\tilde{0}(1)\rangle=(0,1,0,0,0)^{\text{T}}$, $\vert0,2\rangle=(0,0,1,0,0)^{\text{T}}$, $\vert0,1\rangle=(0,0,0,1,0)^{\text{T}}$, and $\vert1,\tilde{1}(1)\rangle=(0,0,0,0,1)^{\text{T}}$, where ``$\text{T}$'' denotes the matrix transpose. Based on the matrices $M_{33}$ and $M_{22}$, we can obtain the eigenvalues of the Hamiltonian $H_{\text{eff}}^{(2)}$ as
\begin{align}
\varepsilon_{0} & =0,\nonumber \\
\varepsilon_{1} & =-\sqrt{\left(\Omega_{1}(t)A_{0,0}^{(1)}\right)^{2}+\left(\Omega_{2}(t)A_{0,2}^{(1)}\right)^{2}}=-\varepsilon_{2},\nonumber \\
\varepsilon_{3} & =-\Omega_{1}(t)A_{1,1}^{(1)}=-\varepsilon_{4},
\end{align}
and the corresponding eigenstates
\begin{align}
\vert\phi_{0}(t)\rangle & =(\Omega_{2}(t)A_{0,2}^{(1)}\vert0,0\rangle -\Omega_{1}(t)A_{0,0}^{(1)}\vert0,2\rangle)/\varepsilon_{2}, \nonumber \\
\vert\phi_{1}(t)\rangle & =\frac{\Omega_{1}(t)A_{0,0}^{(1)}\vert0,0\rangle-\varepsilon_{2}\vert1,\tilde{0}(1)\rangle +\Omega_{2}(t)A_{0,2}^{(1)}|0,2\rangle}{\sqrt{2}\varepsilon_{2}}, \nonumber \\
\vert\phi_{2}(t)\rangle & =\frac{\Omega_{1}(t)A_{0,0}^{(1)}\vert0,0\rangle+\varepsilon_{2}\vert1,\tilde{0}(1)\rangle +\Omega_{2}(t)A_{0,2}^{(1)}|0,2\rangle}{\sqrt{2}\varepsilon_{2}}, \nonumber \\
\vert\phi_{3}(t)\rangle & =(\vert1,\tilde{1}(1)\rangle-\vert0,1\rangle)/\sqrt{2},\nonumber \\
\vert\phi_{4}(t)\rangle & =(\vert1,\tilde{1}(1)\rangle+\vert0,1\rangle)/\sqrt{2}.
\end{align}
Here the state $\vert\phi_{0}(t)\rangle$ is the so-called dark state, which does not include the component of state $\vert1,\tilde{0}(1)\rangle$. If the choices of two pulse driving fields $\Omega_{1,2}(t)$ guarantee adiabatic evolution of $\vert\phi_{0}(t)\rangle$~\cite{gaubatz1990Population,bergmann1998Coherent,vitanov2017Stimulated}, then the population transfer from state $\vert 0,0\rangle$ to $\vert 0,2\rangle$ can be realized in the absence of the dissipation of the system.

To prove the population transfer between the states $\vert 0,0\rangle$ and $\vert 0,2\rangle$, we plot the first Gaussian wave
packets of the two pulse driving fields $\Omega_{1,2}(t)$, the zero-photon state populations $P_{\vert0,m\rangle}=\vert\langle0,m\vert\psi(t)\rangle\vert^{2}$, and the single-photon state populations $P_{\vert1,\tilde{m}(1)\rangle}=\vert\langle1,\tilde{m}(1)\vert\psi(t)\rangle\vert^{2}$ as functions of the time $\omega_{m}t$ in Figs.~\ref{Fig2}(c)-\ref{Fig2}(e). Here we consider that the initial state of the system is $\vert 0,0\rangle$, and the two pulse driving fields $\Omega_{1,2}(t)$ satisfy the condition of STIRAP. It can be seen that the population transfer from state $\vert 0,0\rangle$ to $\vert 0,2\rangle$ is realized in the absence of the dissipation, i.e., when $\omega_{m}t\geq1800$, we have $P_{\vert0,2\rangle}\approx1$ and the populations of the other states are approximately zero.

In the realistic physical system, we need to consider the dissipation of the system. In the presence of dissipation, the dynamics of the system can be described by quantum master equation~\cite{scully1997Quantum}. However, in the ultrastrong-coupling regime, the dynamics of the system can not be described correctly by the standard quantum master equation. Hence, the dynamics of the system in the ultrastrong-coupling regime is governed by the dressed-state master equation~\cite{Hu2015Quantum}
\begin{align}\label{DSME}
\frac{d\rho(t)}{dt}&=i[\rho(t),H_{r}]+\gamma_{m}(n_{\text{th}}+1)\mathcal{D}[b-\beta a^{\dagger}a]\rho(t)\nonumber\\
&\quad+\gamma_{m}n_{\text{th}}\mathcal{D}[b^{\dagger}-\beta a^{\dagger}a]\rho(t)+\kappa\mathcal{D}[a]\rho(t)\nonumber\\
&\quad+4\gamma_{m}(k_{B}T_{b}/\omega_{m})\beta^{2}\mathcal{D}[a^{\dagger}a]\rho(t),
\end{align}
where the Hamiltonian $H_{r}$ is given in Eq.~(\ref{Hamr}), $\kappa$ ($\gamma_{m}$) is the decay rate of the cavity (mechanical) mode. $n_{\text{th}}=[\exp(\hbar\omega_{m}/k_{B}T_{b})-1]^{-1}$ is the thermal phonon occupation number at temperature $T_{b}$, with $k_{B}$ being the Boltzmann constant. The Lindblad superoperators are defined by $\mathcal{D}[o]\rho(t)=[2o\rho(t)o^{\dagger}-\rho(t)o^{\dagger}o-o^{\dagger}o\rho(t)]/2$. For simplicity, we only consider the case of zero temperature, i.e., $T_{b}=0$\,K and $n_{\text{th}}=0$. By numerically solving Eq.~(\ref{DSME}), we can obtain the density operator $\rho(t)$ of the system at time $t$, and then the zero-photon state populations $P_{\vert0,m\rangle}(t)=\text{Tr}[\vert0,m\rangle\langle0,m\vert\rho(t)]$ and the single-photon state populations $P_{\vert1,\tilde{m}(1)\rangle}(t)=\text{Tr}[\vert1,\tilde{m}(1)\rangle\langle1,\tilde{m}(1)\vert\rho(t)]$ can be calculated.
%%%%%%%%%%%%%%%%%%%%%
\begin{figure}
\center
\includegraphics[width=0.46 \textwidth]{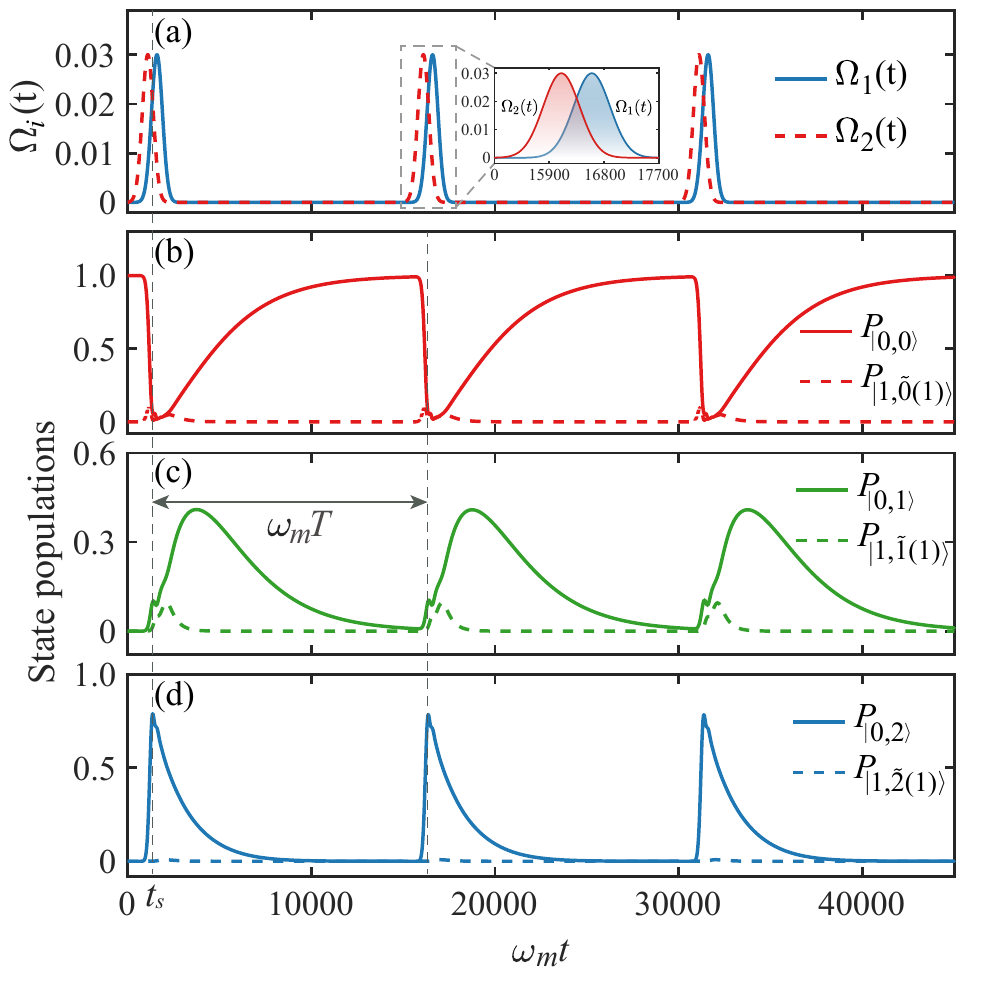}
\caption{(Color online) (a) Two Gaussian pulse driving fields $\Omega_{i}(t)$ $(i=1,2)$ as a function of the time $\omega_{m}t$. (b)-(d) The state populations $P_{\vert0,m\rangle}(t)$ and $P_{\vert1,\tilde{m}(1)\rangle}(t)$ ($m=0,1,2$) as functions of the time $\omega_{m}t$ in the presence of dissipation. Other parameters are $\Omega_{0}/\omega_{m}=0.03$, $\omega_{m}\sigma=300$, $\omega_{m}t_{1}=1600$, $\omega_{m}t_{2}=1100$, $\omega_{m}T=15000$, $g/\omega_{m}\approx0.765$, $\kappa/\omega_{m}=0.002$, $\gamma_{m}/\omega_{m}=0.0004$, $n_{\text{th}}=0$, $\Delta_{1}=-g^{2}/\omega_{m}$, and $\Delta_{2}=-g^{2}/\omega_{m}-2\omega_{m}$.}
\label{Fig3}
\end{figure}
%%%%%%%%%%%%%%%%%%%%%%%

To study the dynamical evolution of the state populations of the system in the presence of dissipation, we plot the state populations $P_{\vert0,m\rangle}(t)$ and $P_{\vert1,\tilde{m}(1)\rangle}(t)$ ($m=0,1,2$) as functions of the time $\omega_{m}t$ at the optomechanical coupling strength $g/\omega_{m}\approx0.765$, as shown in Figs.~\ref{Fig3}(b)-\ref{Fig3}(d). Similarly, we consider an initial state $\vert0,0\rangle$ of the system, i.e., $P_{\vert0,0\rangle}(0)=1$. In Fig.~\ref{Fig3}(a), we plot the two pulse driving fields $\Omega_{1,2}(t)$ as a function of the time $\omega_{m}t$. Here we choose the time interval between the consecutive pulses $\gamma_{m}T\gg1$ such that the system goes back to the initial state $\vert0,0\rangle$ before the arrival of the next Gaussian wave packet. It can be seen from Figs.~\ref{Fig3}(b)-\ref{Fig3}(d) that due to the presence of dissipation, the maximal value of the state population $P_{\vert0,2\rangle}(t)$ at time $t=t_{s}+kT$ is smaller than 1. In addition, due to the decay of the mechanical mode, each phonon in state $\vert0,2\rangle$ is emitted in an intrinsic temporal structure corresponding to the spontaneous emission of the Fock state ~\cite{munoz2014Emitters,strekalov2014Bundle,munoz2018Filtering,bin2020Phonon,bin2021ParitySymmetryProtected,ma2021Antibunched}, the system then goes back to the initial state $\vert0,0\rangle$. The state $\vert0,2\rangle$ is again generated for the next Gaussian pulse, which means that the dynamical super-Rabi oscillation $\vert0,0\rangle\leftrightarrow\vert0,2\rangle$ can be realized under the action of two Gaussian pulse driving fields.
%%%%%%%%%%%%%%%%%%%%%
\begin{figure}
\center
\includegraphics[width=0.46 \textwidth]{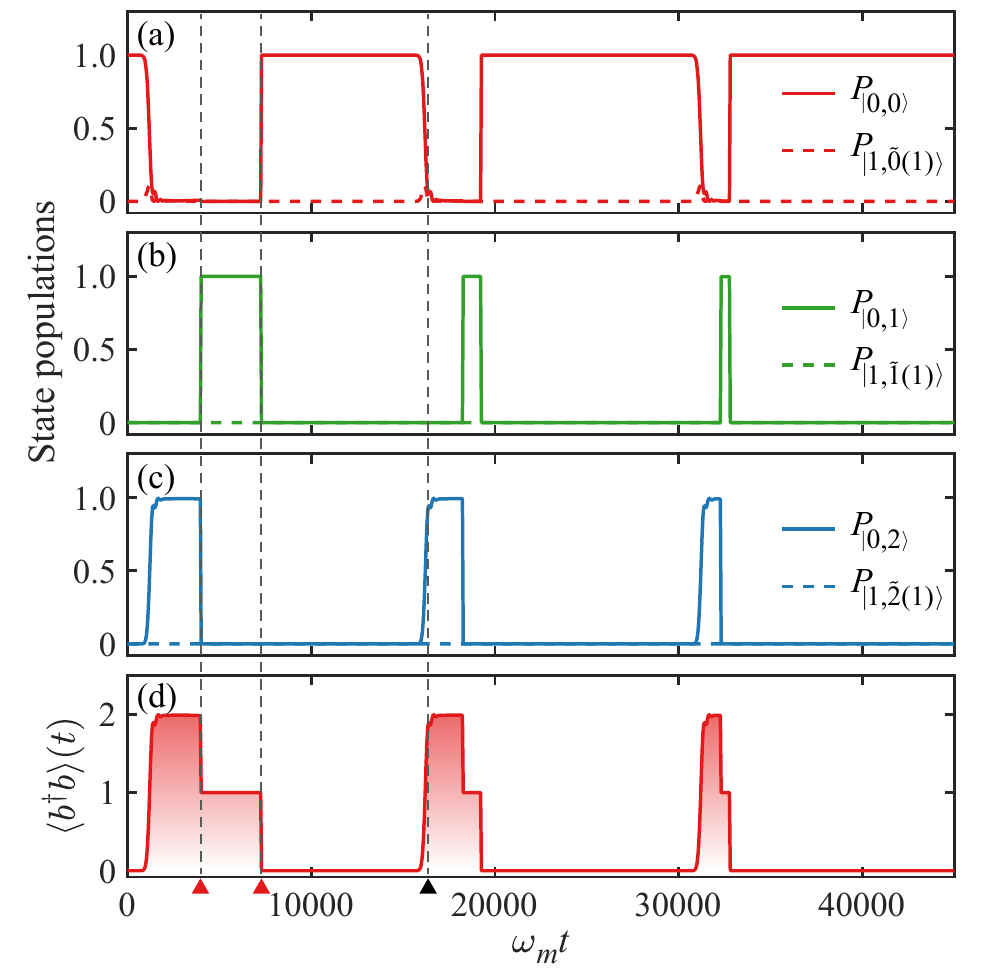}
\caption{(Color online) (a)-(c) Quantum trajectory of the state populations $P_{\vert0,m\rangle}(t)$ and $P_{\vert1,\tilde{m}(1)\rangle}(t)$ ($m=0,1,2$). (d) Quantum trajectory of the average phonon number $\langle b^{\dagger}b \rangle(t)$. Other parameters are $\Omega_{0}/\omega_{m}=0.03$, $\omega_{m}\sigma=300$, $\omega_{m}t_{1}=1600$, $\omega_{m}t_{2}=1100$, $\omega_{m}T=15000$, $g/\omega_{m}\approx0.765$, $\kappa/\omega_{m}=0.002$, $\gamma_{m}/\omega_{m}=0.0004$, $n_{\text{th}}=0$, $\Delta_{1}=-g^{2}/\omega_{m}$, and $\Delta_{2}=-g^{2}/\omega_{m}-2\omega_{m}$.}
\label{Fig4}
\end{figure}
%%%%%%%%%%%%%%%%%%%%%%%

\section{Dynamical emission of phonon pairs\label{tpbesec}}

In this section, we study the dynamical emission of phonon pairs and the statistical properties of the dynamical phonon-pair emission. Concretely, we employ a quantum Monte Carlo approach to simulate individual trajectory of the system. In Figs.~\ref{Fig4}(a)-\ref{Fig4}(c), we show a quantum trajectory of the state populations $P_{\vert0,m\rangle}(t)$ and $P_{\vert1,\tilde{m}(1)\rangle}(t)$ ($m=0,1,2$) at the ratio $g/\omega_{m}\approx0.765$. Based on the technique of STIRAP, the value of the population $P_{\vert0,2\rangle}(t)$ in the state $\vert 0,2\rangle$ at time $t=t_{s}$ is approximately equal to 1. Due to the trigger of the dissipation of the mechanical mode, the first phonon is emitted (indicated by the first red triangle at the bottom of the figure) and the wave function collapses to the one-phonon state $\vert0,1\rangle$ with almost unit probability, as shown in Fig.~\ref{Fig4}(b). Immediately, the second phonon is emitted during the mechanical lifetime (the second red triangle), as shown in Fig.~\ref{Fig4}(a). This means that the strongly correlated phonon pairs are emitted in a very short temporal window and the wave function of the system collapses to the zero-phonon state $\vert0,0\rangle$. After the arrival of the next Gaussian wave packet, the two-phonon state $\vert0,2\rangle$ (the black triangle) is prepared again for the next emission of phonon pairs. Hence, under the action of two Gaussian pulse driving fields, the dynamical emission of phonon pairs can be realized by choosing the appropriate time interval $T$. Figure~\ref{Fig4}(d) shows a quantum trajectory of the average phonon number $\langle b^{\dagger}b \rangle(t)$ at the ratio $g/\omega_{m}\approx0.765$. Here we can see that the dynamical cascade-phonon-emission process $\vert0,2\rangle\rightarrow\vert0,1\rangle\rightarrow\vert0,0\rangle$ occurs in a very short time window.
%Until the emission of the first phonon, the time to maintain with two phonons is $1/(2\gamma_{m})$. After the emission of the first phonon, the single phonon is remained for a time $1/\gamma_{m}$.

\begin{figure}
\center
\includegraphics[width=0.47 \textwidth]{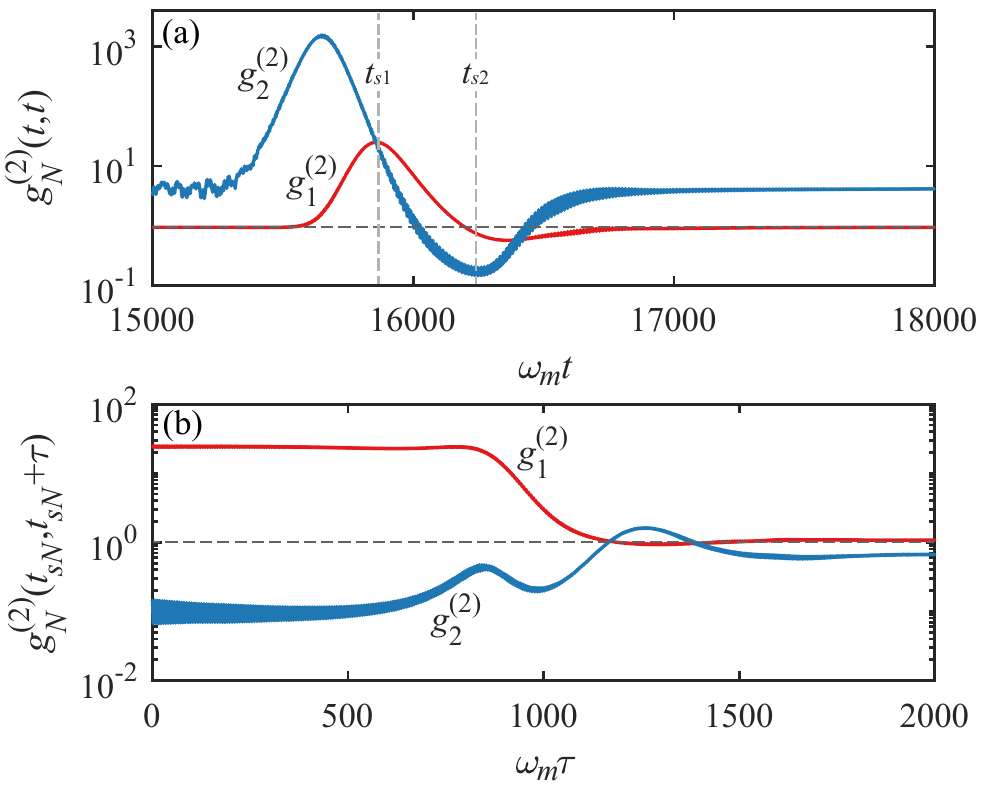}
\caption{(Color online) (a) One period of the equal-time second-order correlation functions $g_{N}^{(2)}(t,t)$ as a function of the time $\omega_{m}t$ with $N=1$ (the red line) and $N=2$ (the blue line). The $t_{s1}$ and $t_{s2}$ correspond to the maximum value of $g_{1}^{(2)}(t,t)$ and the minimum value of $g_{2}^{(2)}(t,t)$, respectively. (c) The time-delay second-order correlation functions $g_{N}^{(2)}(t_{sN},t_{sN}+\tau)$ ($t_{s1}$ and $t_{s2}$ are indicated in the upper panel) with $N=1$ (the red line) and $N=2$ (the blue line). Other parameters are $\Omega_{0}/\omega_{m}=0.03$, $\omega_{m}\sigma=300$, $\omega_{m}t_{1}=1600$, $\omega_{m}t_{2}=1100$, $\omega_{m}T=15000$, $g/\omega_{m}\approx0.765$, $\kappa/\omega_{m}=0.002$, $\gamma_{m}/\omega_{m}=0.0004$, $n_{\text{th}}=0$, $\Delta_{1}=-g^{2}/\omega_{m}$, and $\Delta_{2}=-g^{2}/\omega_{m}-2\omega_{m}$.}
\label{Fig5}
\end{figure}

For the pulse driving fields, we cannot study the steady-state correlation function of the system. Hence, in order to investigate the quantum statistics of the dynamical phonon-pair emission, we numerically calculate the standard and generalized equal-time second-order correlation functions of the mechanical mode~\cite{munoz2014Emitters,bin2020Phonon}
\begin{subequations}
\begin{align}
g_{1}^{(2)}(t,t)&=\frac{\langle b^{\dagger}(t)b^{\dagger}(t)b(t)b(t)\rangle}{\langle b^{\dagger}b(t)\rangle^{2}},\\
g_{2}^{(2)}(t,t)&=\frac{\langle b^{\dagger2}(t)b^{\dagger2}(t)b^{2}(t)b^{2}(t)\rangle}{\langle b^{\dagger2}b^{2}(t)\rangle^{2}}.
\end{align}
\end{subequations}
In Fig.~\ref{Fig5}(a), we display one period of the equal-time second-order correlation functions $g_{N}^{(2)}(t,t)$ ($N=1,2$) as a function of the time $\omega_{m}t$. It can be seen that the value of the standard correlation function $g_{1}^{(2)}(t,t)$ at time $t=t_{s1}$ is maximum and $g_{1}^{(2)}(t_{s1},t_{s1})>1$, which means that super-Poisson of single phonon occurs at time $t=t_{s1}$. Furthermore, we can see that the value of the generalized correlation function $g_{2}^{(2)}(t,t)$ at time $t=t_{s2}$ is minimum and $g_{2}^{(2)}(t_{s2},t_{s2})<1$ corresponding to sub-Poisson of phonon pairs.
%%%%%%%%%%%%%%%%%%%%%
%%%%%%%%%%%%%%%%%%%%%%%

To further characterize the statistical properties of the dynamical phonon-pair emission, we also numerically calculate the standard and generalized time-delay second-order correlation functions of the mechanical mode
\begin{subequations}
\begin{align}
g_{1}^{(2)}(t_{s1},t_{s1}+\tau)&=\frac{G_{1}^{(2)}(t_{s2},t_{s2}+\tau)}{\langle b^{\dagger}b(t_{s1})\rangle\langle b^{\dagger}b(t_{s1}+\tau)\rangle},\\
g_{2}^{(2)}(t_{s2},t_{s2}+\tau)&=\frac{G_{2}^{(2)}(t_{s2},t_{s2}+\tau)}{\langle b^{\dagger2}b^{2}(t_{s2})\rangle\langle b^{\dagger2}b^{2}(t_{s2}+\tau)\rangle}.
\end{align}
\end{subequations}
where $G_{1}^{(2)}(t_{s1},t_{s1}+\tau)=\langle b^{\dagger}(t_{s1})b^{\dagger}(t_{s1}+\tau)b(t_{s1}+\tau)b(t_{s1})\rangle$ and $G_{2}^{(2)}(t_{s2},t_{s2}+\tau)=\langle b^{\dagger2}(t_{s2})b^{\dagger2}(t_{s2}+\tau)b^{2}(t_{s2}+\tau)b^{2}(t_{s2})\rangle$. Figure~\ref{Fig5}(b) shows the time-delay second-order correlation functions $g_{N}^{(2)}(t_{sN},t_{sN}+\tau)$ for $N=1,2$, where $\tau$ is the time delay. Here $t_{s1}$ and $t_{s2}$ correspond to, respectively, the times of the maximum value in $g_{1}^{(2)}(t,t)$ and the minimum value in $g_{2}^{(2)}(t,t)$ in Fig.~\ref{Fig5}(a). As shown in Fig.~\ref{Fig5}(b), the numerical result shows that $g_{1}^{(2)}(t_{s1},t_{s1})>g_{1}^{(2)}(t_{s1},t_{s1}+\tau)$ and $g_{2}^{(2)}(t_{s2},t_{s2})<g_{2}^{(2)}(t_{s2},t_{s2}+\tau)$ are satisfied. This means that the two phonons contained in each phonon pair are bunched, but the relation between phonon pairs and  phonon pairs is antibunched, that is, the system behaves as an antibunched phonon-pair emitter.

\section{Discussion and Conclusion \label{conclusion}}
We present some discussions on the experimental parameters in this theoretical scheme. To implement the present physical scheme, the key point is to realize the ultrastrong optomechanical coupling strength. Currently, the strong optomechanical coupling strength (i.e., $g/2\pi=1.6$\,MHz) has been realized in a superconducting circiut~\cite{Pirkkalainen2015Cavity}. In particular, it has been estimated in Ref.~\cite{Pirkkalainen2015Cavity} that a coupling strength up to $g/2\pi=100$\,MHz is in principle accessible with an optimized device. In our simulations, we use the following parameter: $g/\omega_{m}\approx0.765$, $\kappa/\omega_{m}=0.002$, $\gamma_{m}/\omega_{m}=0.0004$, and $\Omega_{0}/\omega_{m}=0.03$ (e.g., $\omega_{m}/2\pi=100$\,MHz, $g/2\pi=76.5$\,MHz, $\kappa/2\pi=0.2$\,MHz, $\gamma_{m}/2\pi=0.04$\,MHz, and $\Omega_{0}/2\pi=3$\,MHz). We want to point out that these parameters are experimentally accessible in a superconducting circuit, but there still exists some challenges for current experimental technology.

In conclusion, we have proposed an efficient scheme to realize the dynamical phonon-pair emission in a cavity optomechanical system, where the optical cavity is driven by two Gaussian pulse driving fields. In the absence of the dissipation of the system, the population transfer from $\vert0,0\rangle$ to $\vert0,2\rangle$ can be realized based on the technique of STIRAP. By studying the quantum trajectories of the state populations and the average phonon number in the system, we found that the dynamical emission of phonon pairs can be observed under appropriate parameter conditions. Particularly, by numerically calculating the standard and generalized second-order correlation functions of the mechanical mode, we found that the cavity optomechanical system can behave as an antibunched phonon-pair emitter when the time interval between the consecutive pulses $\gamma_{m}T\gg1$. Our proposal can also be extended to achieve an antibunched $n$-phonon emitter, which has potential applications in quantum information processing
and quantum metrology.

\begin{acknowledgments}
%This work is supported by the National Natural Science Foundation of China (Grants No.~12074030 and No.~U1930402). J.-Q. L. is supported in part by National Natural Science Foundation of China (Grants No.~11774087, No.~11822501, No.~12175061, and No.~11935006), Hunan Science and Technology Plan Project (Grant No.~2017XK2018), and the Science and Technology Innovation Program of Human of Hunan Province (Grants No.~2020RC4047 and No.~2021RC4029). F.Z. is supported in part by the National Natural Science Foundation of China (Grants No.~12147109) and the China Postdoctoral Science Foundation (Grant No.~2021M700360).
This work is supported in part by the National Natural Science Foundation of China (Grants No.~12074030, No.~U1930402, No.~11774087, No.~11822501, No.~12175061, No.~11935006, and No.~12147109), Hunan Science and Technology Plan Project (Grant No.~2017XK2018), the Science and Technology Innovation Program of Human of Hunan Province (Grants No.~2020RC4047 and No.~2021RC4029), and the China Postdoctoral Science Foundation (Grant No.~2021M700360).
\end{acknowledgments}

\end{document}